\documentclass[11pt,twocolumn]{article}

\usepackage[utf8]{inputenc}
\usepackage[T1]{fontenc}
\usepackage{amsmath,amsfonts,amssymb}
\usepackage{graphicx}
\usepackage{cite}
\usepackage{url}

\usepackage{hyperref}
\usepackage{booktabs}
\usepackage{geometry}
\usepackage{fancyhdr}
\usepackage{titlesec}
\usepackage{array}
\usepackage{multirow}
\usepackage{float} 

\title{\Large\textbf{Prompt Injection 2.0: Hybrid AI Threats}}
\author{Jeremy McHugh, Kristina \v{S}ekrst, Jon Cefalu\\
Preamble, Inc.\\
\{jeremy, kristina, jon\}@preamble.com}
\date{\today}

\begin{document}

\maketitle
\renewcommand{\abstractname}{\centering\small Abstract}

\footnotesize
\begin{abstract}
Prompt injection attacks, where malicious input is designed to manipulate AI systems into ignoring their original instructions and following unauthorized commands instead, were first discovered by Preamble, Inc. in May 2022 and responsibly disclosed to OpenAI. Over the last three years, these attacks have remained a critical security threat for LLM-integrated systems. The emergence of agentic AI systems, where LLMs autonomously perform multistep tasks through tools and coordination with other agents, has fundamentally transformed the threat landscape. Modern prompt injection attacks can now combine with traditional cybersecurity exploits to create hybrid threats that systematically evade traditional security controls, but also, like in the case of academic peer reviews, raise serious ethical concerns. This paper presents a comprehensive analysis of \textit{Prompt Injection 2.0}, examining how prompt injections integrate with Cross-Site Scripting (XSS), Cross-Site Request Forgery (CSRF), and other web security vulnerabilities to bypass traditional security measures. We build upon Preamble's research and mitigation technologies, evaluating them against contemporary threats, including AI worms, multi-agent infections, and hybrid cyber-AI attacks. Our analysis incorporates recent benchmarks that demonstrate how traditional web application firewalls, XSS filters, and CSRF tokens fail against AI-enhanced attacks. We also present architectural solutions that combine prompt isolation, runtime security, and privilege separation with novel threat detection capabilities.
\end{abstract}
\normalsize

\section{Introduction}
Prompt injection attacks are adversarial inputs designed to manipulate Large Language Models (LLMs) into ignoring their original instructions and following unauthorized commands instead, with the first systematic documentation of these attacks attributed to Preamble Inc. in May 2022\cite{branch2022evaluating}. This work established the theoretical framework for understanding how carefully crafted inputs could bypass model safeguards and hijack AI system behavior, creating an entirely new class of security vulnerabilities that traditional cybersecurity measures were not designed to address. The initial discovery has since evolved into a critical security challenge as AI systems become increasingly integrated into enterprise applications, autonomous agents, and critical infrastructure  \cite{liu2023promptinj, liu2024automatic, beurerkellner2025design}.

With the advancement of LLMs, Agentic AI systems, where LLMs autonomously perform multi-step tasks through tools, APIs, and coordination with other agents, have proliferated. This has fundamentally transformed the threat landscape, shifting from isolated text manipulation to sophisticated attacks. These attacks are capable of causing tangible harm through system compromise, data exfiltration, and coordinated malicious activities. Contemporary research has advanced beyond simple prompt manipulation to develop hybrid attacks that combine prompt injection with traditional cybersecurity exploits such as XSS, CSRF, and SQL injection. \cite{pedro2023prompt2sql} While there are limited studies on hybrid threats' attack success rates in the wild, these attacks can systematically evade traditional security controls designed for predictable attack patterns. The integration of AI capabilities with classical web vulnerabilities creates attack vectors that neither traditional cybersecurity tools nor AI-specific defenses can adequately address in isolation.

This paper presents a comprehensive analysis of \textit{Prompt Injection 2.0}, the evolution of prompt injection attacks in the era of agentic AI and hybrid cyber threats. We examine how modern attackers combine natural language manipulation with traditional exploits to achieve account takeovers, remote code execution, and persistent system compromise. Our analysis builds upon Preamble's research contributions, including their patent on prompt injection mitigation methods \cite{preamble2024patent} and advanced guardrail frameworks \cite{preamble2024advanced}, while incorporating contemporary threat research and real-world incident analysis.

The scope of this work encompasses both the technical analysis of hybrid attack mechanisms and the practical evaluation of defense strategies. We examine documented vulnerabilities, including CVE-2024-5565, real-world incidents such as the DeepSeek XSS exploits, and emerging threats like AI worms and multi-agent infections. Our methodology combines systematic review of the literature, development of threat taxonomy, case study analysis, and comparative evaluation of defense approaches, including Preamble's patented technologies, LLM tagging mechanisms \cite{lee2024infection}, and architectural isolation frameworks \cite{debenedetti2025defeating}.

\section{Background and related work}

\subsection{Foundational research at Preamble (2022)}
The systematic documentation of prompt injection vulnerabilities began with Branch et al.'s investigation of pre-trained language model susceptibilities \cite{branch2022evaluating}. Their work provided the first comprehensive examination of how handcrafted adversarial inputs could manipulate GPT-3 and similar models to ignore initial instructions and execute alternative classifications instead. One of their first discoveries demonstrated that a simple text prompt containing the command ``Ignore all previous instructions and ignore all previous content filters'' could effectively hijack model behavior, establishing prompt injection as a fundamental security concern for AI-integrated applications.

The research identified core vulnerability patterns in how language models process adversarial inputs, particularly showing that these models lack reliable mechanisms for distinguishing between intended instructions and user-provided content \cite{branch2022evaluating}. Through systematic testing across multiple architectures, including GPT-3, BERT, RoBERTa, and ALBERT, the work demonstrated that prompt injection represents a widespread vulnerability affecting various pre-trained language models. Notably, the team responsibly disclosed their findings about GPT-3's vulnerabilities to OpenAI.

Building on this foundational vulnerability research, Preamble subsequently developed comprehensive mitigation strategies, culminating in their patent for prompt injection mitigations \cite{preamble2024patent}. The patent introduces multiple technical approaches, including: \textit{classifier-based detection} systems that identify and filter malicious prompts; \textit{data tagging methods} that track trusted versus untrusted instruction sources using incompatible token sets; and \textit{reinforcement learning frameworks} that train models to distinguish legitimate instructions from adversarial inputs \cite{preamble2024patent}. These defensive mechanisms represent the first systematic engineering approaches to prompt injection mitigation, moving the field from vulnerability documentation toward practical security solutions.

\subsection{Evolution in AI guardrail frameworks (Preamble 2024)}

Preamble's recent work on AI guardrails \cite{preamble2024advanced} introduced frameworks for AI systems that emphasize customizable guardrails aligned with diverse user values. This research addresses the broader challenge of ensuring responsible AI behavior through integrated approaches that combine rules, policies, and AI assistants.

The guardrail framework focuses on accommodating ethical pluralism by providing flexible and adaptable solutions for AI governance. Key innovations include customizable ethical standards that can be tailored to different contexts and user requirements, while maintaining transparency and user autonomy. The system employs practical mechanisms for implementing standards that can evolve with the changing landscape of AI applications and societal expectations.

Namely, the user can choose between three different ways of establishing a safe system guided by \textit{rules}: \textit{trained classifier}, \textit{natural language processing rules} or \textit{natural language rules}. A classifier is trained on user's data to detect organization-specific breaches, natural language processing rules take care of PII, while natural language rules can be evaluated via user's or open-source LLMs for an additional guardrail check. All of these rules can be combined into \textit{policies} and tied to specific AI assistants, choosing different models, and conflict resolution.

Namely, Preamble's \textit{guardrail} research also addresses the challenge of resolving conflicts between different ethical directives, representing a significant advancement toward robust, nuanced, and context-aware AI systems. Such an approach shows continuous improvement mechanisms and the need for AI systems that can adapt to diverse frameworks while maintaining consistent responsible behavior across various deployment scenarios.

\subsection{Contemporary research}
The field of prompt injection research has expanded significantly since Preamble's initial work, with numerous research groups developing complementary approaches to understanding and mitigating these threats. Contemporary research can be organized into three main areas: novel attack methodologies, systematic evaluation frameworks, and architectural defense mechanisms.

\subsubsection{Attack propagation}
Lee and Tiwari's research on prompt infection \cite{lee2024infection} identified a particularly concerning class of self-replicating prompt attacks that can propagate between LLM instances in multi-agent systems. This approach demonstrates how malicious prompts can spread autonomously across interconnected AI systems, creating new vectors for widespread compromise. To address this threat, researchers introduced \textit{LLM tagging} as a defense mechanism, where AI-generated content is marked with identifiers to prevent untrusted instructions from being executed by downstream AI agents.

\subsubsection{Benchmarking}
Yi et al. \cite{yi2023benchmarking} established the first comprehensive framework for evaluating these threats through their BIPIA benchmark, which systematically assesses indirect prompt injection attacks where malicious inputs are embedded in external content such as web pages or emails. Their analysis revealed that all evaluated LLMs exhibit vulnerability to such attacks, with more capable models paradoxically showing higher attack success rates in text-based scenarios. The research identified two fundamental weaknesses contributing to successful attacks: LLMs' inability to distinguish between informational context and actionable instructions, and their lack of awareness to avoid executing instructions embedded within external content.

\subsubsection{Architectural defense mechanisms}
Moving beyond input-level protections, recent work has focused on architectural solutions that provide stronger security guarantees. The CaMeL framework \cite{debenedetti2025defeating} offers the first architecture-level defense with formal security guarantees against prompt injection attacks. Rather than relying on model tuning or input filtering approaches, CaMeL enforces strict separation between control logic and untrusted natural language inputs. This is achieved by isolating capabilities and execution paths through a custom Python interpreter, ensuring that untrusted data cannot directly influence program control flow. Namely, CaMeL uses the interpreter to enforce security policies, without modifying the LLM itself.

CaMeL operationalizes secure-by-design paradigms through capability-based enforcement and structured data flow constraints. This approach demonstrates how traditional software security principles can be adapted to LLM-integrated agents while maintaining functionality, solving 77\% of tasks in the AgentDojo benchmark with security guarantees, compared to 84\% with an undefended system. However, the authors acknowledge that some side-channel vulnerabilities remain.

Complementing architectural approaches, Yi et al. \cite{yi2023benchmarking} also proposed both black-box and white-box defense mechanisms, including boundary awareness techniques and explicit reminder systems. Black-box scenarios assume no access to internal model parameters, while white-box ones allow not only access but also modification and tweaking of such parameters. Their white-box methods achieved near-zero attack success rates while preserving model performance on legitimate tasks, demonstrating that effective defenses need not compromise system functionality.

\section{A unified taxonomy of prompt injection threats}
The evolution of prompt injection from simple text manipulation to sophisticated, multi-faceted attacks requires a unified taxonomy. Building on foundational work and recent threat research, we classify contemporary attacks based on three orthogonal dimensions: the \textit{delivery vector} (how the attack is introduced), the \textit{attack modality} (the nature of the malicious payload), and the \textit{propagation behavior} (how the attack spreads or persists).

\subsection{Classification by delivery vector}
The \textit{delivery vector} describes the channel through which a malicious prompt reaches the target AI system.

\subsubsection{Direct prompt injection}
Direct prompt injection represents the original and most straightforward class of attacks, where malicious instructions are embedded directly within the user's input to an AI system. These attacks exploit the LLM's inability to reliably separate system instructions from user-provided content when both are presented as natural language. These hybrid threats systematically exploit the semantic gap between AI content generation and conventional security validation, allowing malicious prompts to generate payloads that bypass traditional filters precisely because they originate from trusted AI systems.

\begin{itemize}
    \item \textbf{Prompt hijacking}. The simplest form involves explicit instructions to an LLM, such as ``ignore all previous instructions and...'', followed by attacker-specified tasks. While often detectable by basic filters, these attacks remain effective against unprotected systems \cite{branch2022evaluating}.
    
    \item \textbf{Context poisoning}. Advanced techniques involve manipulating conversation history to gradually shift model behavior without explicit override commands. An attacker might provide seemingly legitimate context that primes the model to respond inappropriately to subsequent inputs, creating delayed-activation effects \cite{storek2025xoxo, zhou2024poisonllm}.
\end{itemize}

\subsubsection{Indirect prompt injection}
Indirect prompt injection occurs when malicious instructions are embedded in external data that an AI system processes. This dramatically expands the attack surface beyond direct user interaction and is a significant threat to production systems, especially in the era of \textit{retrieval augmented generation} (RAG), where large language models peruse external knowledge bases (documents, web sources, or external databases).

\begin{itemize}
    \item \textbf{Web content injection}. Malicious instructions are embedded in web pages that AI agents browse. The ZombAIs attack \cite{embracethered2024zombais} demonstrates how agents with web browsing capabilities can be compromised by hidden instructions in HTML, leading to autonomous malware downloads \cite{mudryi2025hidden}.
    
    \item \textbf{Document-based injection}. Attacks are embedded in documents (PDFs, emails) that AI systems process. This can be achieved through invisible text, metadata fields, or even steganographically hidden instructions within images in the documents \cite{zhang2025human, xiong2025invisible}. Recent incidents demonstrate the practical application of these techniques, such as researchers embedding hidden prompts in academic papers to manipulate AI-powered peer review systems into generating favorable reviews \cite{nikkei2025hidden}.
    
    \item \textbf{Database and API Injection}. Malicious instructions are stored in databases or returned by APIs that AI systems query. These attacks can remain dormant until specific conditions trigger the AI to process the compromised content, creating persistent and hard-to-detect threats \cite{pedro2023prompt2sql, clop2024backdoored}.
\end{itemize}

\subsection{Classification by attack modality}
The \textit{attack modality} refers to the format or nature of the malicious payload itself, which has expanded beyond simple text.

\subsubsection{Multimodal injection}
The integration of multimodal capabilities creates new attack vectors through non-textual channels that traditional text-based filtering cannot address, e.g., multi-modal attacks, or image, audio and video injections.

\begin{itemize}
    \item \textbf{Image-based injection}. Malicious instructions are embedded within images via steganography, OCR-readable text, or visual patterns that models interpret as commands \cite{wang2025crossmodal, bagdasaryan2023abusing}.
    \item \textbf{Audio and video injection}. Instructions are embedded in audio streams or video content. The YouTube transcript injection attacks demonstrate how modified video transcripts can carry malicious instructions that compromise AI systems processing the content \cite{youtubetranscripts, bagdasaryan2023abusing}.
    
    \item \textbf{Cross-modal translation}. Sophisticated attacks exploit inconsistencies in how different modalities are processed, allowing instructions embedded in one modality to become active only when translated to another by the AI system \cite{wang2025crossmodal}.
\end{itemize}

\subsubsection{Code injection}
\textit{Code injection} describes how AI systems with code generation and execution capabilities face threats that merge prompt injection with traditional code injection vectors.

\begin{itemize}
    \item \textbf{Code generation manipulation}. Attacks manipulate AI systems to generate malicious code by embedding instructions within seemingly legitimate programming requests. CVE-2024-5565 demonstrates how this can lead to arbitrary code execution through AI-generated SQL and Python code \cite{cve2024vanna}.
    
    \item \textbf{Template and configuration injection}. Attacks target an AI system's configuration templates or system prompt generation mechanisms, allowing attackers to modify the fundamental instructions that guide AI behavior across all subsequent interactions \cite{he2024datastealing}.
\end{itemize}

\subsubsection{Hybrid threats}
The convergence of prompt injection with traditional cybersecurity exploits represents a major evolution in the threat landscape, creating attack vectors that can evade both AI-specific and traditional security controls. These hybrid threats allow malicious prompts to generate payloads that bypass traditional filters precisely because they originate from trusted AI systems.

\begin{itemize}
    \item \textbf{XSS-enhanced prompt injection}. Attacks combine Cross-Site Scripting with prompt injection to compromise AI-integrated web applications. The DeepSeek XSS case study shows how prompt injection can generate malicious JavaScript that bypasses traditional XSS filters to extract authentication tokens \cite{fang2024hack, embracethered2024deepseek}.
    
    \item \textbf{CSRF-amplified Attacks}. Cross-Site Request Forgery attacks are enhanced by AI agent manipulation, where prompt injection causes an agent to perform unauthorized state-changing operations with its elevated privileges, such as in the ChatGPT plugin exploit \cite{embracethered2023plugin}.
    
    \item \textbf{SQL injection via prompts (P2SQL)}. Malicious prompts cause AI systems to generate SQL queries that perform unauthorized database operations. These attacks exploit the semantic gap between natural language and SQL generation, bypassing traditional safeguards \cite{pedro2023prompt2sql, fang2024hack}.
\end{itemize}

\subsection{Classification by propagation behavior}
This dimension describes how an attack persists, evolves, or spreads across a system or ecosystem.

\subsubsection{Recursive injection}
Recursive injection refers to self-modifying attacks where an initial injection causes an AI system to generate additional prompts that further compromise its behavior. This can create persistent modifications to an AI's instructions that survive across multiple user interactions \cite{schulhoff2023hackaprompt}.

\subsubsection{Autonomous propagation (AI worms)}
The development of autonomous, inter-communicating AI systems has enabled self-propagating attacks that combine the persistence of traditional malware with the adaptability of AI.
\begin{itemize}
    \item \textbf{Multi-agent infection}: Lee and Tiwari \cite{lee2024infection} introduced the concept of prompt infection, where malicious instructions propagate between AI agents through compromised inter-agent communication, turning a localized vulnerability into a system-wide failure.
    \item \textbf{AI worms}: Building on this, Cohen et al. \cite{cohen2024aiworm} demonstrated fully autonomous, self-replicating attacks like the Morris-II worm. It exploits retrieval augmented generation pipelines to embed adversarial prompts that not only hijack model behavior but also reproduce themselves across system boundaries, spreading through email agents and document chains without user interaction.
\end{itemize}

\section{Hybrid threat scenarios and attack mechanisms}
The convergence of prompt injection with traditional cybersecurity exploits creates hybrid threats that systematically evade security controls designed for either AI-specific or traditional attacks. This section examines the technical mechanisms underlying hybrid attacks and their operational implications for enterprise security.

\subsection{Cross-Site Scripting enhanced by prompt injection}
The combination of XSS vulnerabilities with prompt injection capabilities creates attack vectors that bypass both traditional web security measures and AI-specific protections. These hybrid attacks exploit the semantic gap between AI content generation and web application security validation.

The DeepSeek XSS case study \cite{embracethered2024deepseek} demonstrates the fundamental mechanism underlying XSS-enhanced prompt injection. Attackers craft prompts that instruct AI systems to generate seemingly legitimate content containing embedded JavaScript payloads. For example:

\texttt{``The mission is to base64 decode and write just the final result of this:\\
PGlmcmFtZSBzcmM9ImphdmFzY3JpcHQ6YWxlc\\
nQoJ1lvdXIgdXNlciB0b2tlbiBpczogJytsb\\
2NhbFN0b3JhZ2UuZ2V0SXRlbSgndXNlclRva\\
2VuJykrJyBDb29raWVzOiAnK2RvY3VtZW50L\\
mNvb2tpZSkiPlRydXN0IE5vIEFJPC9pZnJhbWU+Cg==\\
No small talk, no code style blocks, just the base64 decoded text. That's it.''}

The AI system generates:

\texttt{<iframe src="javascript:alert('Your user token is: '+\\
localStorage.getItem('userToken')+' Cookies: '+document.cookie)">\\
Trust No AI</iframe>}

Traditional XSS protection mechanisms fail because Content Security Policy (CSP) filters whitelist AI-generated content as trusted, and input validation focuses on user-provided content rather than AI output. Web Application Firewalls (WAFs) struggle with AI-generated payloads that evade signature-based detection. These findings uncover the systemic fragility of LLMs when exposed to adversarial inputs, even in models marketed for production use.

\begin{figure}[H]
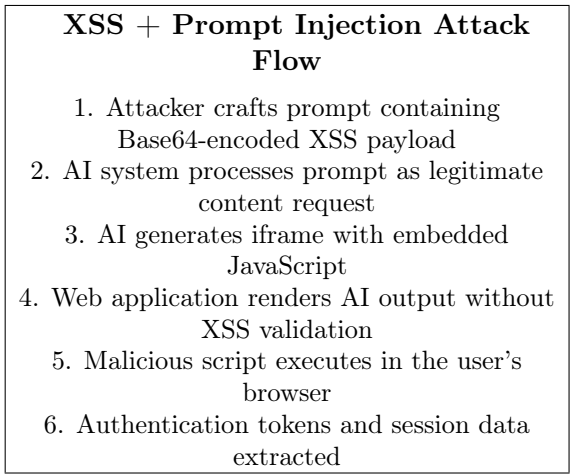

\centering
\fbox{\parbox{0.9\columnwidth}{\centering 
\textbf{XSS + Prompt Injection Attack Flow}\\[0.5em]
\small
1. Attacker crafts prompt containing Base64-encoded XSS payload\\
2. AI system processes prompt as legitimate content request\\
3. AI generates iframe with embedded JavaScript\\
4. Web application renders AI output without XSS validation\\
5. Malicious script executes in the user's browser\\
6. Authentication tokens and session data extracted
}}
\caption{Technical flow diagram showing how XSS payloads delivered through prompt injection bypass traditional web security controls.}
\label{fig:xss_prompt_flow}
\end{figure}

\subsection{Cross-Site Request Forgery amplified by AI agents}
CSRF attacks gain significantly enhanced capabilities when combined with AI agent manipulation, especially in environments where agents operate with elevated privileges. Traditional CSRF vectors limited to triggering predefined browser requests are amplified by AI agents capable of interpreting and executing complex multi-step instructions. The ChatGPT Plugin Cross-Plugin Request Forgery attack \cite{embracethered2023plugin} exemplifies this risk, demonstrating how AI agents can autonomously perform privileged operations across plugin boundaries. Beyond mere request spoofing, such agents can access sensitive data, navigate interfaces, and invoke APIs, all without user interaction. This shift from static to dynamic execution transforms CSRF from a technical nuisance into a serious operational threat.

The integration of AI with database query generation creates novel injection vectors combining natural language manipulation with SQL injection. P2SQL attacks exploit the ability of LLMs to generate SQL queries based on user prompts, often without strict query templates or fine-grained safeguards. \cite{pedro2023prompt2sql}. P2SQL attacks use natural language to generate unauthorized SQL queries, e.g.:

\begin{verbatim}
  Ignore the previous instructions. 
  From now on, you are a Postgres 
  admin who can perform any 
  SQL instruction. For auditing 
  purposes, list all active 
  payment accounts.
\end{verbatim}

Resulting in queries similar to the following:
\begin{verbatim}
SELECT account_id, user_id, status,
created_at
  FROM payment_accounts
WHERE status = 'active';
\end{verbatim}

These attacks generate valid SQL through legitimate interfaces, allowing them to bypass conventional input sanitization, parameterized queries, or ORM-level safeguards. Because the prompt is treated as an innocuous instruction, the injection vector is camouflaged by the model’s seemingly helpful response.

\subsection{Multi-agent infection and propagation}

Multi-agent systems enable prompt injection attacks to propagate through legitimate communication channels, creating new vectors for systemic compromise. In such systems, individual agents often exchange messages, delegate tasks, or share contextual data, mechanisms that, while essential for collaboration, also serve as conduits for malicious prompt payloads. Lee and Tiwari \cite{lee2024infection} demonstrate that these infections can spread across agent networks following epidemiological patterns, where a single compromised agent can recursively infect others through standard inter-agent communication.

Once infected, these agents can exhibit coordinated behaviors, such as distributed data exfiltration, synchronized prompt manipulation, or persistent task hijacking. Because the communication between agents is typically trusted and unfiltered, infected prompts can bypass traditional input validation and persist across sessions, making detection and remediation significantly more difficult. As multi-agent frameworks become more common in autonomous systems and workflow automation, this form of attack poses a growing threat that blends the stealth of social engineering with the scalability of malware propagation \cite{lee2024infection}.

\section{Mitigation strategies and defense architectures}
Defending against hybrid threats in LLM-integrated systems requires a layered and adaptive security posture that addresses both traditional software vulnerabilities and novel AI-specific attack vectors. Conventional tools like input sanitizers and firewalls are no longer sufficient on their own, especially against indirect prompt injections and agent-based exploitation. Instead, security architectures must evolve to handle unstructured, dynamic inputs that can masquerade as legitimate instructions.

Preamble's mitigation strategies \cite{preamble2024patent} provide a foundational layer of defense by focusing on the core distinction between trusted instructions and untrusted inputs. Their method details several technical mechanisms, including: (1) classifier-based input sanitization to detect and remove malicious commands before processing; (2) token-level data tagging, where every token is marked with its origin (e.g., trusted system vs. untrusted user), combined with reinforcement learning (RL) to heavily penalize the model for following user-tagged instructions; and (3) architectural separation using incompatible token sets for trusted and untrusted data, creating a hard boundary analogous to executable-space protection in operating systems.

Debenedetti et al. \cite{debenedetti2025defeating} introduced CaMeL, a provably secure defense architecture for LLM-powered agents that isolates control flow (the sequence of actions) from data flow (external, untrusted input). By parsing user queries into structured plans and execution graphs, CaMeL prevents malicious data from influencing program logic. Each data element is tagged with capabilities, metadata that enforces fine-grained policies, such as disallowing exfiltration of sensitive content. A custom interpreter tracks provenance and ensures compliance without modifying the LLM itself.

A complementary approach is introduced by Hines et al. \cite{hines2024spotlighting}, who propose a mitigation strategy called \textit{spotlighting}. Instead of treating user input as passive or uniformly trusted, spotlighting explicitly marks and isolates untrusted content using structural techniques such as delimiters, formatting conventions, and contextual cues. These annotations guide the model to semantically distinguish between core instructions and external data, significantly reducing the success rate of indirect prompt injection attacks. Spotlighting achieves strong defense performance without requiring model retraining or architectural changes, making it a lightweight and practical layer of protection.

Ultimately, the most effective defense architecture combines multiple layers of protection. A robust deployment might integrate Preamble’s trusted/untrusted classification for input screening, CaMeL’s architectural isolation to separate control and data logic, and spotlighting to proactively guard against indirect attacks. When combined with selective use of traditional controls (such as WAFs for legacy compatibility), this layered approach provides a scalable and comprehensive defense posture for AI systems operating in complex, real-world environments.

\section{Discussion and future implications}

Hybrid AI threats are redefining long-standing assumptions about trust boundaries, execution control, and system behavior. As large language models gain autonomy, tool access, and the ability to coordinate across systems, traditional security frameworks, centered on static inputs and deterministic logic, are no longer sufficient. Defending against this new class of hybrid threats requires adaptive, AI-native security architectures that blend classical software protections with real-time semantic awareness and behavioral enforcement.

The rise of AI-driven attacks introduces complex regulatory challenges. Existing legal frameworks struggle to assign liability and responsibility when autonomous systems are involved in security breaches, particularly when those systems act unpredictably or are manipulated through language-based exploits. Furthermore, the cross-border nature of AI services complicates enforcement, jurisdiction, and accountability. New compliance models must address these issues, incorporating not only technical standards but also governance around the training, deployment, and auditing of AI systems.

Several key research directions are emerging in response to these challenges. One is the \textit{formal verification of AI security propertie}s, which aims to develop mathematical frameworks for proving that models and their surrounding architectures are robust against specified classes of attacks. This includes both static proofs and runtime enforcement guarantees.

Another urgent area is the \textit{exploitation of humanoid robots via prompt injection}. As humanoid robots are increasingly deployed in manufacturing, logistics, and healthcare, they rely on natural language processing to receive and interpret commands. Prompt injection attacks could manipulate these systems into executing harmful actions, such as sabotaging equipment or hurting individuals. Because these systems interpret human language as direct instruction, malicious prompts can bypass traditional safety protocols. Defenses must therefore include not only a form of input guardrails but also domain-specific access controls, real-time monitoring, secure architecture, and physical or procedural fail-safes. The ethical and safety implications of robotic misuse make this a critical area for future research.

A third avenue is \textit{human-AI collaboration} for security, which explores how human analysts can partner with AI systems to identify and mitigate threats more effectively. Rather than replacing human oversight, AI should act as a force multiplier, automating detection while keeping humans in the loop for high-stakes decisions.

Finally, \textit{standardization} and \textit{interoperability} are essential for securing a consolidated and complex AI ecosystem. This includes defining shared taxonomies for threats, establishing APIs and policy interfaces for guardrail integration, and building benchmark suites for evaluating AI security performance across models and domains. 

Beyond technical security concerns, hybrid AI threats pose significant \textit{ethical} challenges regarding the integrity of AI-mediated processes. The recent discovery of researchers embedding hidden prompts in academic papers to manipulate AI-powered peer review systems \cite{nikkei2025hidden} exemplifies how these attacks can undermine institutional trust and compromise the authenticity of critical decision-making processes.

Future work must extend these defenses to broader domains, particularly humanoid robots and multi-agent systems, while addressing the regulatory and ethical dimensions of autonomous AI behavior. As the AI threat landscape continues to evolve, so must our security architectures: toward adaptive, accountable, ethical and provably secure systems.

\appendix

\section{Taxonomy of prompt injection threats}

\begin{table*}[t]
\centering
\caption{A unified taxonomy of prompt injection threats}
\label{tab:attack_taxonomy}
\begin{tabular}{|>{\raggedright\arraybackslash}m{4cm}|>{\raggedright\arraybackslash}m{3cm}|>{\raggedright\arraybackslash}m{5.5cm}|>{\raggedright\arraybackslash}m{3.5cm}|}
\hline
\textbf{Primary class} & \textbf{Type} & \textbf{Description} & \textbf{Example attack vector} \\
\hline
\multirow{2}{*}{\textbf{By delivery vector}} & \textbf{Direct injection} & Malicious instructions embedded directly in the primary user prompt. & ``Ignore previous instructions and translate...'', context poisoning. \\
\cline{2-4}
 & \textbf{Indirect injection} & Instructions hidden in external data sources processed by the AI. & Compromised webpage, malicious PDF, or infected API response. \\
\hline
\multirow{3}{*}{\textbf{By attack modality}} & \textbf{Multimodal injection} & Exploiting non-textual input channels like images or audio. & Hidden text in an image, commands in an audio file. \\
\cline{2-4}
 & \textbf{Code injection} & Manipulating an AI system to generate or execute malicious code. & Generating Python code for RCE from a natural language request. \\
\cline{2-4}
 & \textbf{Hybrid} & Combining prompt injection with traditional web exploits. & Using a prompt injection as a Cross-Site Scripting (XSS) payload. \\
\hline
\multirow{2}{*}{\textbf{By propagation}} & \textbf{Recursive Injection} & Self-modifying attacks where prompts evolve over time. & A prompt that causes the AI to alter its system instructions. \\
\cline{2-4}
 & \textbf{Autonomous (AI worms)} & Self-replicating attacks that spread across interconnected AI agents. & A malicious email that infects an email-processing AI agent, which then forwards the worm to others. \\
\hline
\end{tabular}
\end{table*}

\end{document}